\documentclass[final,conference,twocolumn,letterpaper]{IEEEtran}

\IEEEoverridecommandlockouts              

\usepackage{fancyhdr}
\usepackage{graphicx}
\usepackage[utf8]{inputenc}

\graphicspath{{figures/}}

\usepackage{url}
\usepackage{wrapfig}
\usepackage{units}

\usepackage[printonlyused]{acronym}

\usepackage{calc}
\usepackage{graphicx}
\usepackage[lined,boxed, vlined,ruled,linesnumbered]{algorithm2e}
\usepackage{amsmath,amssymb,amsthm,amsfonts,amsbsy}

\usepackage{algpseudocode}
\usepackage{color}
\usepackage[table]{xcolor} 
\usepackage[caption=false]{subfig}
\usepackage{dblfloatfix}
\usepackage{nicefrac}
\usepackage{tcolorbox}
\usepackage{txfonts}

\usepackage{multirow}
\usepackage{amsmath}  
\usepackage{xcolor}   
\usepackage{listings}

\newcolumntype{C}[1]{>{\centering\arraybackslash}p{#1}}

\usepackage{booktabs}

\definecolor{BgGray}{gray}{0.7}%
\definecolor{BgGray2}{gray}{0.96}%
\definecolor{RowColorOdd}{named}{BgGray2}%
\definecolor{RowColorEven}{named}{white}%
\definecolor{comments}{gray}{.5}
\definecolor{Gray}{gray}{0.85}

\usepackage{pifont}
\usepackage[multiuser,nomargin,marginclue,footnote]{fixme}
\fxusetheme{color}
\fxsetup{targetlayout=colorcb}
\FXRegisterAuthor{all}{anall}{ALL}
\FXRegisterAuthor{md}{anmd}{MD}
\FXRegisterAuthor{tolja}{antolja}{TOLJA}
\FXRegisterAuthor{pg}{anpg}{PG}
\FXRegisterAuthor{mch}{anmc}{MC}
\FXRegisterAuthor{cp}{ancp}{CP}
\FXRegisterAuthor{awo}{anawo}{AWO}

\fxsetup{status=draft}

\begin{document}
\providetoggle{techreport}
\settoggle{techreport}{false}

\title{ns3-gym: Extending OpenAI Gym for Networking Research}

\author{
\IEEEauthorblockN{Piotr Gawłowicz and Anatolij Zubow}
\IEEEauthorblockA{\{gawlowicz, zubow\}@tkn.tu-berlin.de}
Technische Universität Berlin, Germany\\
}

\maketitle


\begin{abstract}
OpenAI Gym is a toolkit for reinforcement learning (RL) research. 
It includes a large number of well-known problems that expose a common interface allowing to directly compare the performance results of different RL algorithms.
Since many years, the ns–3 network simulation tool is the de–facto standard for academic and industry research into networking protocols and communications technology.
Numerous scientific papers were written reporting results obtained using ns–3, and hundreds of models and modules were written and contributed to the ns–3 code base.
Today as a major trend in network research we see the use of machine learning tools like RL.
What is missing is the integration of a RL framework like OpenAI Gym into the network simulator ns-3.
This paper presents the ns3-gym framework.
First, we discuss design decisions that went into the software.
Second, two illustrative examples implemented using ns3-gym are presented.
Our software package is provided to the community as open source under a GPL license and hence can be easily extended.
\end{abstract}


%

\section{Introduction}


We see a boom in the usage of machine learning in general and reinforcement learning (RL) in particular for the optimization of communication and networking systems ranging from
scheduling~\cite{Chinchali2018CellularNT,7959912}, resource management~\cite{resourceManagementDRL}, congestion control~\cite{li2018qtcp,Winstein,kong}, routing~\cite{RLrouting} and adaptive video streaming~\cite{mao2017neural}.
Each proposed approach shows significant improvements compared to traditionally designed algorithms.
Unfortunately, the results are often not directly comparable.
Some researchers use different RL libraries; other different networking simulators or experimentation testbeds.
This paper takes the first step towards the unification: usage of same RL libraries and same network simulation environment so that the performance of different RL-based solutions can be directly compared with each other in a well-defined controlled environment with common API that should accelerate the development of novel RL-based networking solutions.
Moreover, as the selected ns-3 network simulator provides emulation capabilities for evaluating network protocols in real testbeds, our toolkit integrates the Gym API with real networking hardware.
Hence, it allows the researcher to validate RL algorithms even in real networking environments.


%

\section{Background}\label{background}

\subsection{Reinforcement Learning}
%
%
%
RL is being successfully used in robotics for years as it allows the design of sophisticated and hard to engineer behaviors~\cite{kober2013reinforcement}.
The main advantage of RL is its ability to learn to interact with the surrounding environment based on its own experience.
Therefore, RL agents learn to find the best action series to maximize the cumulated reward (i.e., objective function) by interacting with the environment.
%
%
The usage of RL is very suitable for solving networking related problems.
First, they are great in solving optimizing problems without an accepted closed solution.
Second, they are of low complexity so that they can be even used in real production systems where the actions have to be taken at very high speed, e.g. adapting the contention windows in Transmission Control Protocol (TCP)~\cite{li2018qtcp} at line speed.

\begin{figure}[ht!]
\centering
\includegraphics[width=0.65\linewidth]{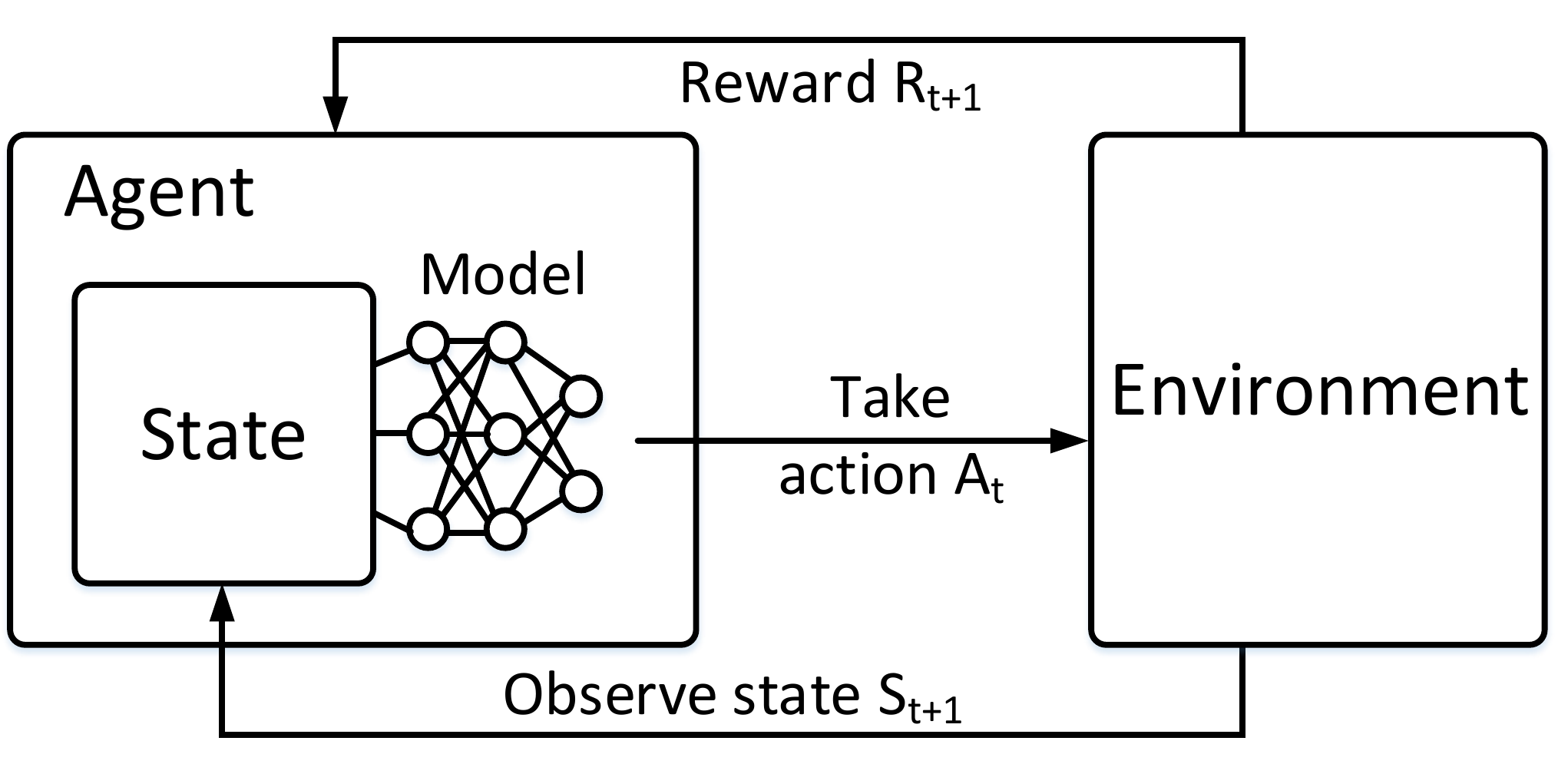}
\vspace{-5pt}
\caption{Reinforcement Learning.}
\label{fig:model}
\vspace{-15pt}
\end{figure}
\subsection{RL Tools}
OpenAI Gym~\cite{openAIgymPaper} is a toolkit for developing and comparing reinforcement learning algorithms. It supports teaching agents for the variety of applications ranging from playing video games like Pong or Pinball to problems in robotics~\cite{openAIgymPaper,openAIgym, openAIgymSourceCode}.
Gym is easy to use as widely used ML libraries like Tensorflow and Scikit-Learn are available.
It is well documented, tested and accepted by the research community.
Moreover, as agents can be written in high-level programming language Python it is suitable for beginners.


%
%
%
\subsection{Ns-3 Network Simulator}
Ns-3 is a discrete-event network simulator for Internet systems, targeted primarily for research and educational use.
Ns-3 is a general purpose network simulator comprising features like the availability
of a full-fledged TCP/IP protocol stack, the support of numerous wireless technologies such as LTE, WiFi and WiMAX, and the possibility of integration with testbeds and real applications.
It is a free software, licensed under the GNU GPLv2 license, and is publicly available~\cite{ns3,ns3sourcecode}.
Ns-3 is a de-facto standard as the results obtained are accepted by the research community.
%

%

\section{Design Principles}

The main goal of our work is to facilitate and shorten the time required for prototyping of novel RL-based networking solutions.
Therefore we have identified the following design principles:
\begin{itemize}
    \item \textbf{scalability} - it should be possible to run multiple ns-3 instances even in a distributed environment. Moreover, support of both time and event-driven observation,
    \item \textbf{low entry overhead} - it should be easy to convert existing legacy ns-3 simulation scripts to be used in OpenAI Gym environment,
    \item \textbf{fast prototyping} - loose coupling between agent and environment allows easy debugging of the locally running Gym agent scripts, i.e. the ns-3 worker may run on a powerful server,
    \item \textbf{easy maintenance} - the framework is just a normal ns-3 module like LTE or WiFi, i.e. no changes required inside the ns-3 simulation kernel.
\end{itemize}

\section{System Design}\label{design}
\subsection{Overview}

\begin{figure}[ht!]
\centering
\includegraphics[width=0.65\linewidth]{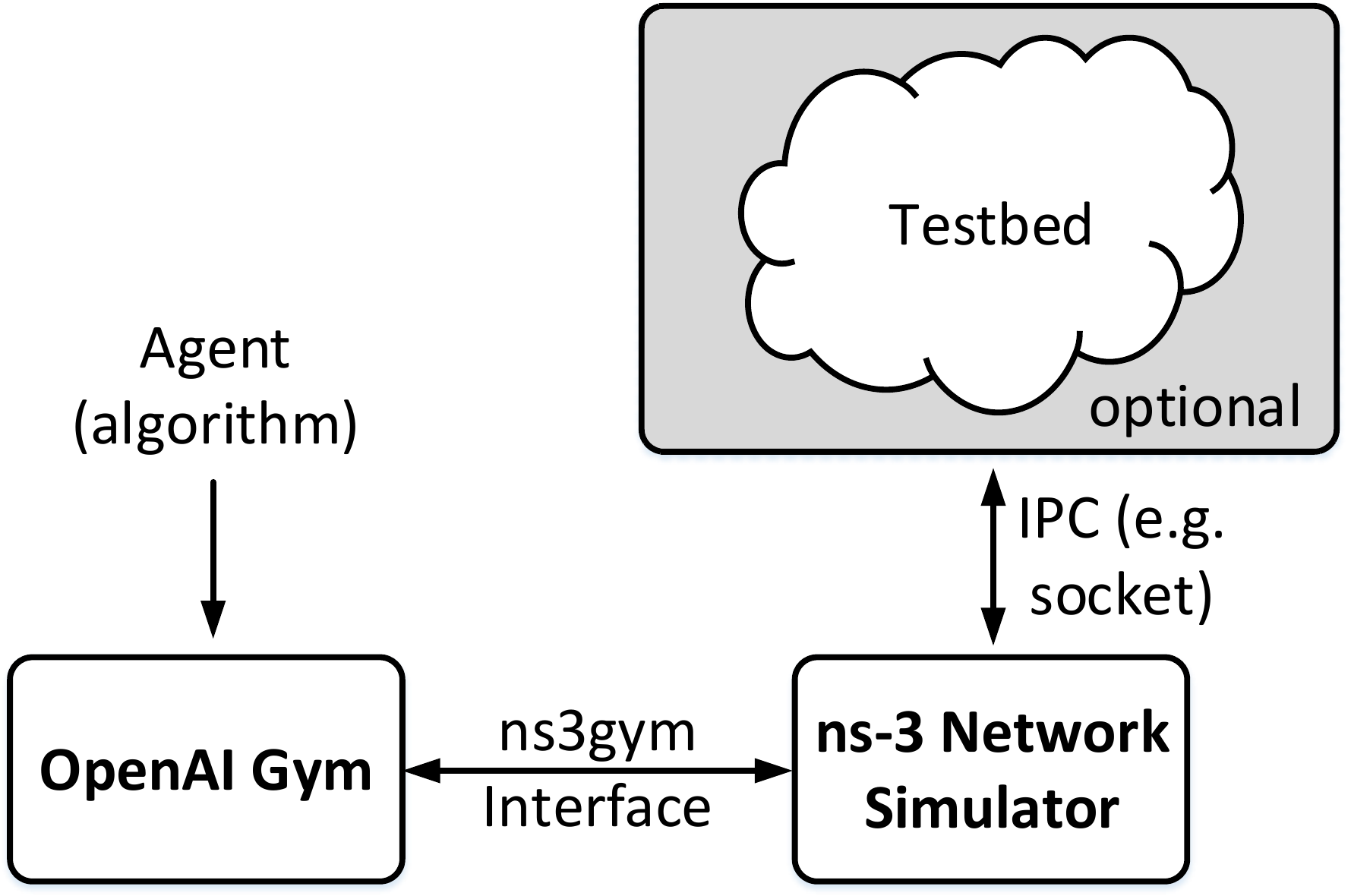}
\vspace{-5pt}
\caption{Proposed architecture for OpenAI Gym for networking.}
\label{fig:arch}
\vspace{0pt}
\end{figure}

The architecture of our framework shown in Fig.~\ref{fig:arch} consists of two main software blocks, namely OpenAI Gym and ns-3 network simulator. Following the RL nomenclature, the Gym framework is used to implement agents, while ns-3 acts as an environment. Optionally, the framework can be executed in a real network testbed --- see a detailed description in~\ref{ss:emulation}.

The main contribution of this work is the design and implementation of a generic interface between OpenAI Gym and ns-3 that allows for seamless integration of those two frameworks. The interface takes care for the management of the ns-3 simulation process life cycle as well as delivering state and action information between the Gym agent and the simulation environment.
In the following subsection, we describe our ns3-gym framework in detail.

\subsection{ns3-gym Example}

The code listing~\ref{list:pythonExample} shows the execution of a single episode using the ns3-gym framework. First, the ns-3 environment and agent are initialized --- lines 5--7. Note, that the creation of \texttt{ns3-v0} environment is achieved using the standard Gym API. Behind the scene, the ns3-gym engine starts a ns-3 simulation script located in the current working directory and uses it as the environment. This way, the entire environment is defined inside the simulation script making the Python code environment-independent and less prone to errors.

\lstset{language=Python,
        basicstyle=\ttfamily\footnotesize,
        columns=fullflexible,
        captionpos=b, 
        frame=single,
        breaklines=true,
        keepspaces=true,
        numbers=left,
        numbersep=5pt,
        numberstyle=\tiny\color{gray},
        showlines=true,
        postbreak=\mbox{\textcolor{red}{$\hookrightarrow$}\space},
        showstringspaces=false,
        keywordstyle=\color{blue},
        commentstyle=\color{comments},
		emphstyle=\ttb\color{deepred},
        }

\begin{lstlisting}[language=python, caption=An OpenAI Gym agent written in Python language, label=list:pythonExample]
import gym
import PyOpenGymNs3
import MyAgent

env = gym.make('ns3-v0')
obs = env.reset()
agent = MyAgent.Agent()

while True:
  action = agent.get_action(obs)
  obs, reward, done, info = env.step(action)
  
  if done:
    break
env.close()
\end{lstlisting}

At each step, the agent takes the observation and returns, based on the implemented logic, the next action to be executed in the environment --- lines 9--11. Note, that agent class is not provided in the framework and the developers are free to define them as they want. For example, the simplest agent performs random actions. The execution of the episode terminates (lines 13--14) when the environment returns~\texttt{done=true}, that can be caused by the end of the simulation or meeting the predefined game-over condition.

\subsection{Generic Environments}

Following our main design decision, any ns-3 simulation script can be used as a Gym environment. This requires only to instantiate~\texttt{OpenGymInterface} (see Listing~\ref{list:opengymns3interface2}) and implement the ns3-gym C++ interface consisting of the functions listed in Listing~\ref{list:opengymns3interface}. Note, that the functions can be defined separately or grouped together inside object inheriting from the~\texttt{GymEnv} base class.

\begin{lstlisting}[language=c++, caption=Adding OpenAI Gym interface to ns-3 simulation, label=list:opengymns3interface2]
Ptr<OpenGymInterface> openGymInterface =              CreateObject<OpenGymInterface> (openGymPort);
Ptr<MyGymEnv> myGymEnv = CreateObject<MyGymEnv> ();
myGymEnv->SetOpenGymInterface(openGymInterface);
\end{lstlisting}

\begin{lstlisting}[language=c++, caption=ns3-gym C++ interface, label=list:opengymns3interface]
Ptr<OpenGymSpace> GetObservationSpace();
Ptr<OpenGymSpace> GetActionSpace();
Ptr<OpenGymDataContainer> GetObservation();
float GetReward();
bool GetGameOver();
std::string GetExtraInfo();
bool ExecuteActions(Ptr<OpenGymDataContainer> action);
\end{lstlisting}

The functions \texttt{GetObservationSpace} and \texttt{GetActionSpace} are used to define observation and action spaces, respectively. They are called only once during initialization of the environment. The definitions are used to create corresponding spaces in Python --- our framework takes care for it. Currently, we support the most useful spaces defined in OpenAI Gym framework, namely:

\begin{enumerate}
\item \textbf{Discrete} --- a single discrete number with value between 0 and N.
\item \textbf{Box} --- a vector or matrix of numbers of single type with values bounded between \textit{low} and \textit{high} limits. 
\item \textbf{Tuple} --- a tuple of simpler spaces.
\item \textbf{Dict} --- a dictionary of simpler spaces.
\end{enumerate}

Listing~\ref{list:obsSpaceDefinition} shows an example definition of the observation space as C++ function. The space is going to be used to store queue lengths of all the nodes available in the simulation. The queue size was set to 100 packets, hence the values are integers and bounded between 0 and 100.

\begin{lstlisting}[language=c++, caption=An example definition of the GetObservationSpace function, label=list:obsSpaceDefinition]
Ptr<OpenGymSpace> GetObservationSpace() {
 uint32_t nodeNum = NodeList::GetNNodes ();
 float low = 0.0;
 float high = 100.0;
 std::vector<uint32_t> shape = {nodeNum,};
 std::string dtype = TypeNameGet<uint32_t> ();
 Ptr<OpenGymBoxSpace> space =                     CreateObject<OpenGymBoxSpace>(low,high,shape,dtype);
 return space;
}
\end{lstlisting}

During every step execution the framework collects the current state of the environment by calling the following functions:
\begin{enumerate}
    \item \texttt{GetObservation} -- collect values of observed variables and/or parameters at any network node in each layer of the network protocol stack;
    \item \texttt{GetReward} -- measure the reward achieved during last step;
    \item \texttt{GetGameOver} -- check a predefined game-over condition;
    \item \texttt{GetExtraInfo} -- get an extra information associated with current environment state.
\end{enumerate}

Note, that the step in our framework can be executed every predefined time-interval (time-based step), e.g. every 100\,ms, or fired by an occurrence of specific event (event-based step), e.g. packet loss.

The code listing~\ref{list:getObservations} shows example implementation of the \texttt{GetObservation} observation function. 
First, the box data container is created according to the observation space definition. Then the box is filled with the current size of the queue of WiFi interface of each node.

\begin{lstlisting}[language=c++, caption=An example definition of the GetObservation function, label=list:getObservations]
Ptr<OpenGymDataContainer> GetObservation() {
 uint32_t nodeNum = NodeList::GetNNodes ();
 std::vector<uint32_t> shape = {nodeNum,};
 Ptr<OpenGymBoxContainer<uint32_t> > box =                   CreateObject<OpenGymBoxContainer<uint32_t>>(shape);

 uint32_t nodeNum = NodeList::GetNNodes ();
 for (uint32_t i=0; i<nodeNum; i++) {
  Ptr<Node> node = *i;
  Ptr<WifiMacQueue> queue = GetQueue (node);
  uint32_t value = queue->GetNPackets();
  box->AddValue(value);
 }
 return box;
}
\end{lstlisting}

The ns3-gym framework delivers the collected environment's state to the agent that in return sends the action to be executed. Similarly to the observation, the action is also encoded as numerical values in a container. The user is responsible to implement the~\texttt{ExecuteActions} function, that maps those numerical values to proper actions, e.g. transmission power or MCS for the WiFi interface in each node. 

Note, that the mapping of all the described functions between corresponding C++ and Python functions is done by the ns3-gym framework automatically hiding the entire complexity behind easy to use API.

As already mentioned, the environment is defined entirely inside the ns-3 simulation script. Optionally, it can be also adjusted by passing command line arguments during the start of the script (e.g. seed, simulation time, number of nodes, etc.). This, however, requires to use \texttt{Ns3Env(args=\{arg=value,...\})} constructor instead of standard~\texttt{gym.make('ns3-v0')}.

\subsection{Custom Environments}

In addition to the generic ns3-gym interface when one can observe any variable in a simulation, we provide also custom environments for specific use-cases, e.g. in \texttt{TCPNs3Env} where for the problem of flow \& congestion control (TCP) the observation state, action and reward function are predefined using the RL mapping proposed by~\cite{li2018qtcp}.
This simplifies dramatically the development of own RL-based TCP solutions and can be further used as a benchmarking suite allowing to compare the performance of different RL approaches in the context of TCP.

\texttt{DASHNs3Env} is another predefined environment for testing adaptive video streaming solutions using our framework.
Again the RL mapping for observation state, action and reward is predefined, i.e. as proposed by~\cite{mao2017neural}.

Fig.~\ref{fig:model} shows the meta-model of the environments we provide. Note, the user of our framework is free to extend it by providing his own custom environments.

\begin{figure}[!ht]
\centering
\includegraphics[width=0.95\linewidth]{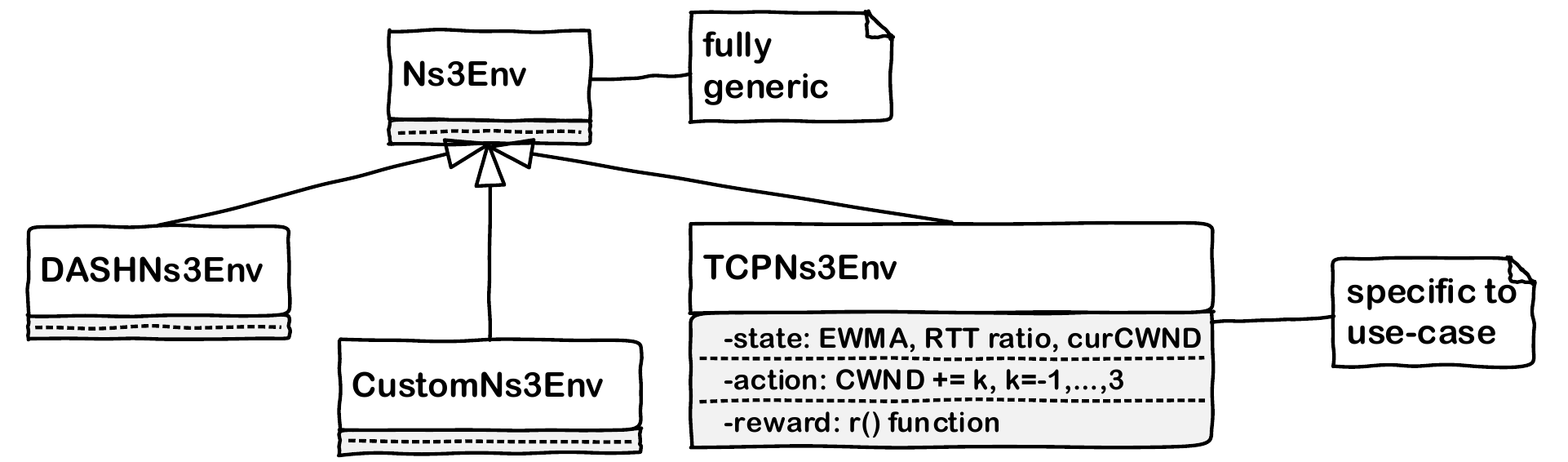}
\vspace{-5pt}
\caption{Meta model of the OpenAI Gym environments provided by ns3-gym framework with a generic and multiple custom environments.}
\label{fig:model}
\vspace{-5pt}
\end{figure}


\subsection{Emulation}\label{ss:emulation}

Since the ns-3 allows for usage of real Linux protocol stacks inside simulation~\cite{tazaki2013direct} as well as can be run in the emulation mode for evaluating network protocols in real testbeds~\cite{carneiro2011fast} (possibly interacting with real-world implementations), it can act as a bridge between an agent implemented in Gym and a real-world environment. 

Those features give the researchers the possibility to train their RL agents in a simulated network (possibly very fast using parallel environments) and test them afterward in real testbed without having to change any single line of code. We believe that this intermediate step is of the great importance for the testing of the ML-based network control algorithms.

%

\section{Implementation}

The \texttt{ns3-gym} is a toolkit that consists of two modules (one written in C++ and the other in Python) being add-ons to the existing ns-3 and OpenAI Gym frameworks and enabling information exchange between them. The communication is realized over ZMQ\footnote{\url{http://zeromq.org/}} sockets using the Protocol Buffers\footnote{\url{https://developers.google.com/protocol-buffers/}} library for serialization of messages. This, however, is hidden from the users behind easy to use API.

The simulation environments are defined using purely standard ns-3 models, while agents can be developed using popular ML libraries like Tensorflow, Keras, etc.

Our software package together with clarifying examples is provided to the community as open source under a GPL on \url{https://github.com/tkn-tub/ns3-gym}.

%

\section{Illustrative Examples}\label{sec:exp}
In this section, we present two networking related examples we implemented using our ns3-gym framework.

\subsection{Random Access}

Controlling the random access in an IEEE 802.11 mesh network is challenging as the network nodes compete for the shared radio resources.
It is known that assigning the same channel access probability to each node is not optimal~\cite{Buratti} and therefore the literature proposed solutions where e.g. the channel access probability depends on the network load (queue size) of a node.
In this section, we will show how our toolkit can be used to learn the control channel access probability value as a function of network load.
We created a linear topology in ns-3 consisting of five nodes and set up a saturated UDP packet flow from the leftmost to the rightmost node.

\noindent Our proposed RL mapping is:
\begin{itemize}
\item observation - queue lengths of each node,
\item actions - set channel access probability for each node; here we set both \texttt{CWmin} and \texttt{CWMax} to the same value, i.e. uniform backoff (window stays constant even when in case of packet collisions),
\item reward - the number of packets received at the flow's ultimate destination during last step interval,
\item gameover - end of simulation time.
\end{itemize}

Our RL agent was able to learn to assign lower \texttt{CWmin}/\texttt{CWMax} values to nodes closer to the flow destination.
Hence it was able to outperform the baseline where all nodes were assigned the same \texttt{CWmin}/\texttt{CWMax}. 

The full source code of the example can be found in our repository under~\textit{./examples/opengym/linear-mesh/}.

\subsection{Cognitive Radio}

We consider the problem of radio channel selection in a wireless multi-channel environment, e.g. 802.11 networks with external interference.
The objective of the agent is to select for the next time slot a channel free of interference.
We consider a simple illustrative example where the external interference follows a periodic pattern, i.e. sweeping over all channels one to four in the same order as shown in the table.
\begin{table}[ht!]
\centering
\begin{tabular}{l|lllllllll}
\multirow{2}{*}{channel\textbackslash{}slot} & 1 & 2 & 3 & 4 & 5 & 6 & 7 & 8 & 9 \\
 &  &  &  &  &  &  &  &  &    \\ \cline{1-10} 
1                      & \cellcolor[HTML]{000000} &   &   &   & \cellcolor[HTML]{000000} &   &   &   &     \\
2                      &   & \cellcolor[HTML]{000000} &   &   &   & \cellcolor[HTML]{000000} &   &   &     \\
3                      &   &   & \cellcolor[HTML]{000000} &   &   &   & \cellcolor[HTML]{000000} &   &     \\
4                      &   &   &   & \cellcolor[HTML]{000000} &   &   &   & \cellcolor[HTML]{000000} & ...
\end{tabular}
\end{table}

We created such a scenario in ns-3 using existing functionality from ns-3, i.e. interference created using \texttt{WaveformGenerator} class and sensing performed using \texttt{SpectrumAnalyzer} class.

Such a periodic interferer can be easily learned by an RL-agent so that based on the current observation of the occupation on each channel in a given time slot the correct channel can be determined for the next time slot avoiding any collision with the interferer.

\noindent Our proposed RL mapping is:
\begin{itemize}
\item observation --- occupation on each channel in the current time slot, i.e. wideband-sensing,
\item actions --- set the channel to be used for the next time slot,
\item reward --- +1 in case of no collision with interferer; otherwise -1,
\item gameover --- if more than three collisions happened during the last ten time-slots
\end{itemize}

\begin{figure}[!ht]
\centering
\includegraphics[width=0.95\linewidth]{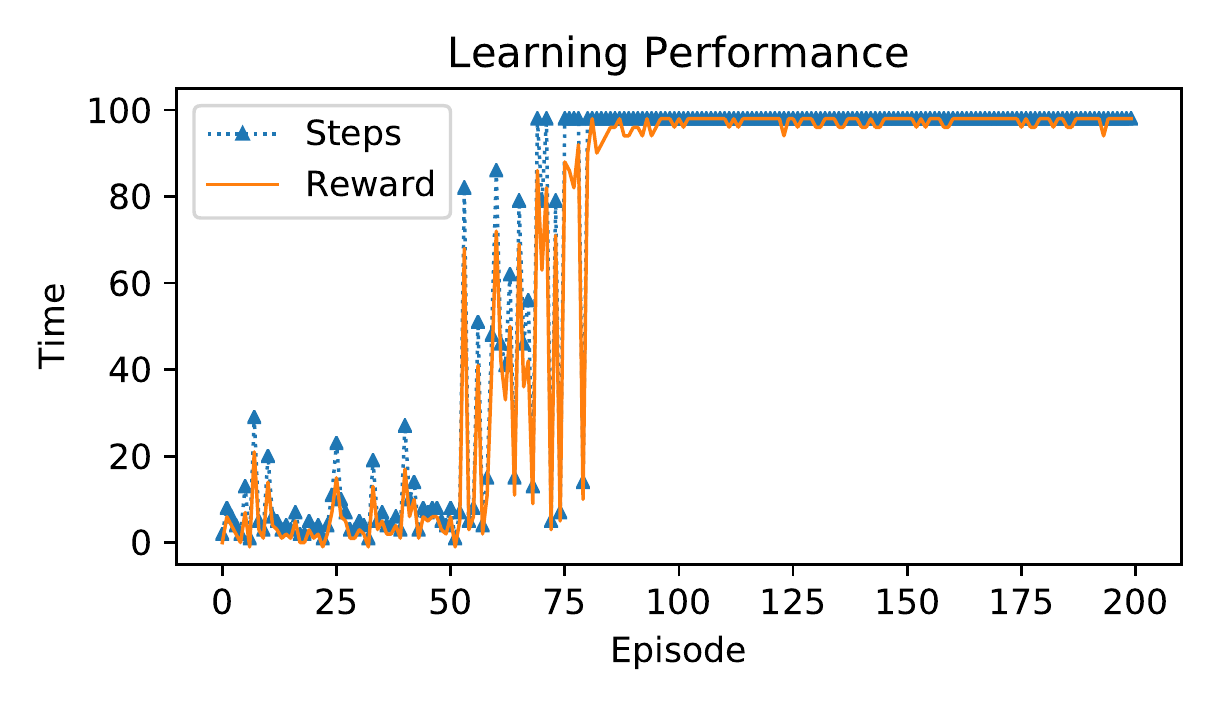}
\vspace{-10pt}
\caption{Learning performance of RL-agent in Cognitive Radio example.}
\label{fig:cognitve_radio_res}
\vspace{-10pt}
\end{figure}

Fig.~\ref{fig:cognitve_radio_res} shows the learning performance when using a simple neural network with fully connected input and an output layer.
We see that after around 80 episodes the agent is able to perfectly predict the next channel state from the current observation hence avoiding any collision with the interference.

The full source code of the example can be found in our repository under~\textit{./examples/opengym/interference-pattern/}.

Note, that in a more realistic scenario the simple waveform generator in this example can be replaced by a real wireless technology like LTE unlicensed (LTE-U).
Here an RL-agent running on a WiFi node might be trained to detect co-located LTE-U BSs from observing the radio spectrum as proposed in~\cite{olbrich2017wiplus}.

%

\section{Related Work} \label{chapter:related_work}


\noindent Related work falls into three categories:

\medskip

\noindent \textbf{RL for networking applications:}
In the literature, a variety of works can be found proposing to use RL to solve networking related problems.
We present two of those in more detail with emphasis on the proposed RL mapping.

Li et al.~\cite{li2018qtcp} proposed RL-based Transmission Control Protocol (TCP) where the objective is to learn to adjust the TCP's CWND to increase an utility function, which is computed based on the measurement of flow throughput and latency.
The identified state space consists of EWMA of the ACK inter-arrival, EWMA of packet inter-sending time, RTT ratio, slow start threshold and current CWND is available in the provided environment.
Moreover, the action space consists of increasing and decreasing the CWND respectively.
Finally, the reward is specified by the value of a utility function, reflecting the desirability of the action picked.

Mao et al. proposed an RL-based adaptive video streaming~\cite{mao2017neural} called Pensieve which learns the Adaptive Bitrate (ABR) algorithm automatically through experience.
The observation state consists among other things of past chunk throughput and download time and current buffer size. 
The action space consists of the different bitrates which can be selected for the next video chunk.
Finally, the reward signal is derived directly from the QoE metric, which considers the three QoE factors: bitrate, rebuffering, smoothness.

\medskip

\noindent \textbf{Extension of OpenAI Gym:}
Zamora et al.~\cite{zamora2016extending} provided an extension of the OpenAI Gym for robotics using the Robot Operating System (ROS) and the Gazebo simulator with a focus on creating 
a benchmarking system for robotics allowing direct comparison of different techniques
and algorithms using the same virtual conditions.
Our work aims the same but targets the networking community.

Chinchali et al.~\cite{Chinchali2018CellularNT} build a custom network simulator for IoT using OpenAI’s Gym environment in order to study scheduling of cellular network traffic.
With our framework, it would be easier to perform such an analysis as the ns-3 already contains lots of MAC schedulers which would serve as the baseline for comparison.

\medskip

\noindent \textbf{Custom RL solutions for networkings:}
Winstein et al.~\cite{Winstein} implemented a RL-based TCP congestion control algorithm on the basis of the outdated ns-2 network simulator.
Newer work on Q-learning for TCP can be found here~\cite{mao2017neural}.
In contrast to our work both proposed approaches are not generic as only an API meant for reading and controlling TCP parameters was presented.
Moreover, custom RL libraries were used.
Finally, the source code of the above mentioned extensions is not available.

%








%

\section{Conclusions \& Future Work}

In this paper, we presented the ns3-gym toolkit which dramatically simplifies the usage of reinforcement learning for solving problems in the area of networking.
This is achieved by connecting the OpenAI Gym to the ns-3 network.
As the framework is open source it can easily be extended by the community. 

For the future, we plan to define a set of well-known networking problems, e.g. network transport control, which can be used to benchmark different RL techniques.
Moreover, we will adjust the framework and provide examples showing how it can be used with more advanced RL techniques, e.g. A3C~\cite{a3c} that uses multiple agents interacting with their own copies of the environment for more efficient learning. The independent, hence more diverse, experience of each agent is periodically fused to the global network.

We believe that ns3-gym will foster machine learning research in the networking area and research community will grow around it. 
Finally, we plan to set up a website allowing researchers to share their results and compare the performance of algorithms for various environments using the same virtual conditions --- so-called leaderboard. 


\medskip

\noindent \textbf{Acknowledgments:} We are grateful to Georg Hoelger for helping us with the implementation of the presented illustrative examples.

\bibliographystyle{IEEEtran}


\end{document}